\documentclass[fleqn,usenatbib,useAMS]{mnras}

\usepackage{graphicx}	
\usepackage{amsmath}	
\usepackage{amssymb}	
\usepackage{multicol}        
\usepackage{bm}		
\usepackage{pdflscape}	

\usepackage[T1]{fontenc}
\usepackage{ae,aecompl}

\usepackage{newtxtext,newtxmath}
\usepackage{subfig}

\title[Lyman Excess H\,{\normalsize \textit{II}} Regions]{A Refined Sample of Lyman Excess H\,{\Large \textbf{II}} Regions}

\author[Marshall \& Kerton]{
Brandon Marshall$^{1}$\thanks{E-mail: bjm@iastate.edu}
and C. R. Kerton$^{1}$
\\
$^{1}$Department of Physics \& Astronomy, Iowa State University, 2323 Osborn Dr., Ames IA 50011, USA\\
}

\date{Accepted XXX. Received YYY; in original form ZZZ}

\pubyear{2018}

\begin{document}
\label{firstpage}
\pagerange{\pageref{firstpage}--\pageref{lastpage}}
\maketitle

\begin{abstract}
A large number (67) of the compact/ultra-compact \ion{H}{ii} regions identified in the CORNISH catalogue were determined to be powered by a Lyman continuum flux in excess of what was expected given their corresponding luminosity. In this study we attempt to reasonably explain away this Lyman excess phenomenon in as many of the 67 $\ion{H}{ii}$ regions as possible through a variety of observational and astrophysical means including new luminosity estimates, new \textit{Herschel} photometry, new distance determinations, the use of different models for dust and ionized gas covering factors, and the use of different stellar calibrations. This phenomenon has been observed before; however, the objects shown to exhibit this behavior in the literature have decidedly different physical properties than the regions in our sample, and thus the origin of the excess is not the same. We find that the excess can be reproduced using OB stellar atmosphere models that have been slightly modified in the extreme ultraviolet. Though the exact mechanism producing the excess is still uncertain, we do find that a scaled up magnetospheric accretion model, often used to explain similar emission from T Tauri stars, is unable to match our observations. Our results suggest that the Lyman excess may be associated with younger \ion{H}{ii} regions, and that it is more commonly found in early B-type stars. Our refined sample of 24 Lyman excess \ion{H}{ii} regions provides an ideal sample for comparative studies with regular \ion{H}{ii} regions, and can act as the basis for the further detailed study of individual regions.
\end{abstract}


\section{Introduction} 
\label{sec:intro}

Massive, OB-type stars are distinguished by their high luminosities, typically >10\textsuperscript{4} \(\textup{L}_\odot\), and extensive Lyman continuum (LyC) flux. In spite of being the focus of much theoretical and observational research, the processes involved in the formation of these stars are still not well established \citep{zinn,Motte}. OB-type stars are often associated with \ion{H}{ii} regions, where the ultraviolet emission from the star has ionized the surrounding interstellar medium (ISM). 

A large number of compact and ultra-compact \ion{H}{ii} regions were observed and catalogued in the Coordinated Radio and Infrared Survey for High-Mass Star Formation \citep[CORNISH;][]{hoare,purcell}. Such regions are of particular interest for studies of massive star formation as they represent a fairly early stage in the evolution of the OB star and the surrounding \ion{H}{ii} region. Intriguingly, a sizable fraction of the \ion{H}{ii} regions in the CORNISH study are apparently being powered by stars that have a LyC photon rate or "flux" ($N_\mathrm{Ly}$) much larger than expected for the observed luminosity of the region \citep[][C15]{cesa15}. The excess $N_\mathrm{Ly}$ is particularly problematic as it is most likely a lower-limit since it is derived from radio observations assuming optically-thin emission and an order unity ionized gas covering factor \citep[e.g.,][]{mats76,ker99}.  These ``Lyman Excess Regions" (LERs) occupy an apparently forbidden region on a plot of $N_\mathrm{Ly}$ versus luminosity (an $N_\mathrm{Ly}$-$L$ plot hereafter; see \autoref{fig:Nly_L_ExcessMyCalcTheirNumbers}).

To understand LERs, we need to either account for an additional source of ionizing photons, or we need to explain why $N_\mathrm{Ly}$ and luminosity have been overestimated or underestimated respectively. There are a number of potential astrophysical origins for the Lyman-excess phenomenon (LEP) including emission from accretion shocks, Balmer continuum ionization, and enhanced LyC emission from the stellar photosphere. Unfortunately there are also a number of more prosaic astrophysical and observational factors (e.g., non-uniform covering factor, incorrect distance, incorrect photometry) that can cause a normal \ion{H}{ii} region appear as a LER. 

Our main goal is to identify a subset of the CORNISH \ion{H}{ii} regions that are bona fide LERs, i.e., regions where we can find no reasonable way to adjust the calculated $N_\mathrm{Ly}$ and luminosity such that they are consistent with values expected for regular \ion{H}{ii} regions. The physical properties (size, luminosity, density) of this refined sample will help us better understand the astrophysics underlying the LEP. The subset will also provide a list of potential targets for individual detailed studies, and act as a high-quality sample that can be used as the basis of comparative studies with other \ion{H}{ii} regions. 

In \autoref{sec:sample} we describe the initial sample of \ion{H}{ii} regions that are potentially LERs. In \autoref{sec:analysis} we investigate the effect of various factors on the LERs classification, and isolate a sample of \ion{H}{ii} regions that can definitely be identified as LERs. We discuss the properties of the refined sample, and explore multiple potential astrophysical origins of the LEP in \autoref{sec:discussion}. Conclusions follow in \autoref{sec:conclusions}.

\section{Sample} \label{sec:sample}

The initial sample consists of the 67 \ion{H}{ii} regions that were identified by C15 as having an excess of LyC photons for a given luminosity, out of the total 200 regions. Each of these sources has been identified in the CORNISH catalogue as being either a compact or ultra-compact \ion{H}{ii} region. The sources have $N_\mathrm{Ly}$ values consistent with main sequence spectral types ranging from approximately B1 to earlier than O6 for both the \citet{panagia} calibration (hereafter PAN73) and the more recent \citet{conti08} calibration. The remaining \ion{H}{ii} regions catalogued in C15 already lie outside of the forbidden region of the $N_\mathrm{Ly}$-$L$ plot, and we do not consider them in detail any further. 

\section{Analysis} \label{sec:analysis}

In the following subsections we describe various ways that the $N_\mathrm{Ly}$ and luminosity values for our sample can be reasonably modified. Some of these actions result in objects that were LERs moving into the allowed region of the  $N_\mathrm{Ly}$-$L$ plot thus removing them from our final sample (see \autoref{tab:distances} and \autoref{tab:photometry}).

\subsection{Luminosity Recalculation} 
\label{sec:recalc}

\begin{figure}	
\includegraphics[width=\columnwidth]{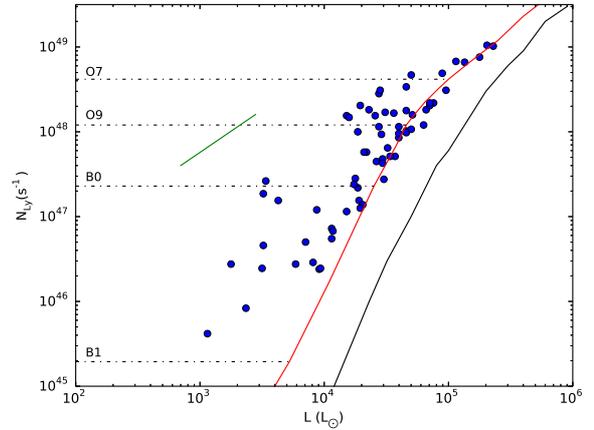}
    \caption{Lyman continuum flux versus the recalculated luminosities using flux densities from C15 for our initial sample of  67 LERs. The red curve corresponds to the $N_\mathrm{Ly}$, log\textsubscript{10}(L/\(\textup{L}_\odot\)) values for single OB-type stars in the \citet{panagia} calibration. For reference, the black dashed lines indicate the $N_\mathrm{Ly}$ for a range of main sequence spectral types. The area between the red and solid black curve is where 90\% of stellar clusters simulated by C15 are located. The green line shows how a source would be displaced if the distance to that source changes by a factor of two.}
    \label{fig:Nly_L_ExcessMyCalcTheirNumbers}
\end{figure}

We calculated the luminosity by numerically integrating each spectral energy distribution (SED) between 21 and 500 $\mu$m  using linear interpolation between the flux densities given in C15. We found that this increased the luminosity for each source by 21\% on average. This was enough to to move 14 sources of the 67 out of the forbidden region as shown in \autoref{fig:Nly_L_ExcessMyCalcTheirNumbers} and summarized in \autoref{tab:distances}. 

\begin{table}
	\centering
	\caption{The number of original LERs, binned in log\textsubscript{10}($N_\mathrm{Ly}$), is shown in the first row. The following rows show how many sources are removed from the forbidden region of the $N_\mathrm{Ly}$-$L$ plot for the given parameter change.}
    \label{tab:distances}
	\begin{tabular}{lcccccc} 
		\hline      
		log\textsubscript{10}($N_\mathrm{Ly}$):  & 46-47 & 47-48 & 48-49 & 49-50 & Total \\
		\hline
		C15 Distance   & 11 & 24 & 28 & 2 & 67\textsuperscript{a}\\
		\hline         
        Recalculated L  & 0 & 4 & 9 & 1 &  14 \\
        \hspace{5mm} $f_c = 0.8$ & 0 & 12 & 11 & 1 & 24 \\
        \hspace{5mm} $f_c = 0.5$  & 6 & 18 & 13 & 2 & 39 \\
        \hspace{5mm} Davies\textsuperscript{b}  & 0 & 8 & 10 & 2 & 20 \\
        \hspace{10mm} $f_c = 0.8$  & 0 & 13 & 12 & 2 & 27  \\
        \hline
        \hline
        MK18 Distance  & 14 & 22 & 28 & 1 &  67\textsuperscript{a} \\        
        \hline
        New Distance Only &  0 & 0 & 2 & 0 & 2 \\
        Recalculated L   & 0 & 6 & 11 & 0 & 17  \\
        \hspace{5mm} $f_c = 0.8$  & 1 & 8 & 14 & 0 & 23 \\ 
        \hspace{5mm} $f_c = 0.5$  & 4 & 14 & 14 & 1 & 33 \\
        \hspace{5mm} Davies  & 0 & 6 & 13 & 1 & 20 \\
        \hspace{10mm} $f_c = 0.8$  & 1 & 10 & 15 & 1 & 27 \\
        \hline
        \multicolumn{6}{p{\linewidth}}{$^\mathrm{a}$Note this value includes two sources with log\textsubscript{10}($N_\mathrm{Ly}$) in the 45-46 range. These are not shown in the table since they remain LERs for all recalculations.}\\
        \multicolumn{6}{p{\linewidth}}{$^\mathrm{b}$ZAMS stellar parameters from \citet{davies}.}\\
		\end{tabular}        
\end{table}

\begin{table}
	\centering
	\caption{As in \autoref{tab:distances}, now using only our new photometry (see \autoref{sec:photometry}).}
	\label{tab:photometry}
	\begin{tabular}{lcccccc} 
		\hline        
		log\textsubscript{10}($N_\mathrm{Ly}$):  & 46-47 & 47-48 & 48-49 & 49-50 & Total \\
		\hline
		C15 Distance   & 6 & 15 & 16 & 2 & 41\textsuperscript{a,b}\\
		\hline         
                New Photometry & 0 & 2 & 7 & 1 & 10 \\
        \hspace{5mm} $f_c = 0.8$  & 0 & 7 & 9 & 1 & 17 \\
        \hspace{5mm} $f_c = 0.5$  & 0 & 9 & 11 & 2 & 22 \\
        \hspace{5mm} Davies\textsuperscript{c}  & 0 & 3 & 7 & 2 & 12 \\
                         \hspace{10mm} $f_c = 0.8$  & 0 & 7 & 9 & 2 & 18 \\
        \hline
        \hline
        MK18 Distance  & 8 & 12 & 18 & 1 &  41\textsuperscript{a,b} \\        
        \hline
               New Photometry  & 0 & 2 & 9 & 0 & 11 \\
         \hspace{5mm} $f_c = 0.8$  & 0 & 4 & 11 & 0 & 15 \\
         \hspace{5mm} $f_c = 0.5$  & 1 & 6 & 13 & 1 & 21 \\
          \hspace{5mm} Davies  & 0 & 3 & 11 & 1 & 15 \\
        \hspace{10mm} $f_c = 0.8$  & 0 & 4 & 12 & 1 & 17  \\
        \hline
        \multicolumn{6}{p{\linewidth}}{$^\mathrm{a}$Note this value includes two sources with log\textsubscript{10}($N_\mathrm{Ly}$) in the 45-46 range. These are not shown in the table since they remain LERs for all recalculations.}\\
        \multicolumn{6}{p{\linewidth}}{$^\mathrm{b}$The total number of sources omits those deemed as medium or low confidence from further consideration.}\\
        \multicolumn{6}{p{\linewidth}}{$^\mathrm{c}$ZAMS stellar parameters from \citet{davies}.}\\       
        \end{tabular}        
\end{table}

\subsection{Distance} 
\label{sec:distance}

We also examined the consequence of changing the distance, which simultaneously affects both calculated estimates of the luminosity of the source as well as the LyC flux (e.g., see the green diagonal line in \autoref{fig:Nly_L_ExcessMyCalcTheirNumbers}). 

Kinematic distances were recalculated using the Bayesian distance calculator of \citet{bays}. This calculator combines a traditional Galactic rotation model with a spiral structure model based on CO observations and trigonometric parallax distances to masers associated with massive star-forming regions. For young objects associated with spiral arms the derived distances are likely superior to those found using a kinematic model with no adjustments made for spiral structure. Radial velocities (V\textsubscript{LSR}) for all of the LERs exist in the literature \citep{anderson2014,urq2013,shirley,anderson2012,anderson2009}. The near-far distance ambiguity was resolved by matching the near-far solution with that of C15. We refer to the recalculated distances as MK18 distances throughout the paper. 

Using the luminosities and $N_\mathrm{Ly}$ from C15, we first only rescaled their values using the MK18 distances, which caused only two sources to move out of the excess region. When using our newly calculated luminosities (see \autoref{sec:recalc}) and the MK18 distances, an additional 15 sources move from the excess to the cluster region. 

\subsection{Photometry} 
\label{sec:photometry}

Obtaining accurate infrared photometry of extended objects in the Galactic plane is challenging. Confusion from unrelated compact and extended sources can be a problem, especially at longer wavelengths due to decreasing resolution, and background estimation can be difficult due to interstellar medium structure. For this reason we  performed our own aperture photometry on each source using the 70, 160, 250, 350, and 500~$\mu$m \textit{Herschel} images (see \autoref{tab:Photometry1}). We further categorized the source photometry into high, medium, and low confidence subcategories, based on source isolation and the potential for confusion when performing the photometry. 

For each source we positioned and adjusted the size of an elliptical aperture manually. An average background was then calculated using an elliptical annulus surrounding the source aperture. At the scale of our sources, fluctuations in the ISM background were minimal, and the background determined in this manner was judged to be appropriate. To explore this more, photometry was done on a subset of the sources using the {\sc{imview}} program \citep{higgs97}. This program creates a background surface (twisted-plane or twisted-quadratic) that is matched to user-selected points surrounding the source of interest. It is ideal for photometry of extended surfaces located on a background gradient or within highly structured background emission.  We found that differences in the individual band flux densities never exceeded $10\%$, and there were no major differences between the luminosities calculated using the two photometry techniques. We recalculated the luminosity of all of the LERs in our initial sample using both C15 and MK18 distances. Error estimates for the luminosities were obtained via a Monte Carlo approach using a conservative $10\%$ error in each of the \textit{Herschel} bands, which when applied after adjusting all other parameters, had no effect in removing additional sources from consideration.  

Of the 67 LERs, we have high confidence in our photometry for 41 of the regions due to their isolation and compact structure. There are 14 regions that were less isolated and have a potential for confusion. These were tagged as medium confidence regions. Finally, 12 regions lay within more complex regions and were categorized as low confidence regions. We removed the 26 sources with medium/low confidence photometry from the LER sample.

We did not make any adjustments to the original $N_\mathrm{Ly}$ values for two reasons. First, as already mentioned in \autoref{sec:intro}, the $N_\mathrm{Ly}$ value is likely a lower limit, and any increase in $N_\mathrm{Ly}$ only increases the Lyman excess. Second, in contrast to the infrared, the sources viewed in the radio continuum are typically very well defined with a uniform background. Thus we are confident that redoing the radio continuum photometry would not have resulted in any significant changes to the derived $N_\mathrm{Ly}$ given by: 
\begin{equation} 
\label{eqn:nly}
N_\mathrm{Ly}\big(\mathrm{s^{-1}}\big) = 9.9\times 10^{43} S_\nu d^2 ,
\end{equation}
where S\textsubscript{$\nu$} is the integrated flux density at 5 GHz in mJy, and $d$ is the distance to the source in kpc.

\begin{figure}	
	\includegraphics[width=\columnwidth]{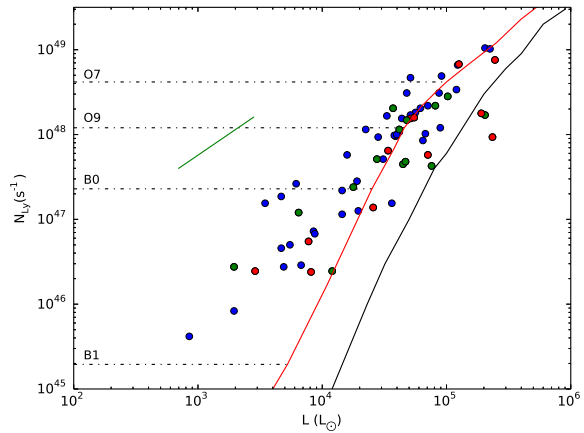}
    \caption{Lyman continuum flux of the 67 LERs, as in \autoref{fig:Nly_L_ExcessMyCalcTheirNumbers}, with luminosities recalculated using the new photometry and C15 distances. The difference in colour for each source corresponds to our confidence level in the photometry: blue is high, green is medium, and red is low confidence.}
    \label{fig:Nly_l_ExcessMyPhot}
\end{figure}

The recalculated luminosities resulted in 10 or 11 of the high confidence photometry LERs moving out of the forbidden region using the C15 or MK18 distances respectively. \autoref{fig:Nly_l_ExcessMyPhot} shows the position of the 67 LERs with luminosities recalculated using the new photometry and the C15 distances. In this case, the LER data points can only shift in luminosity. When the MK18 distances are used, the $N_\mathrm{Ly}$ are also scaled to the appropriate value meaning the original LER data points can shift in both $N_\mathrm{Ly}$ and luminosity (see \autoref{fig:Nly_l_MyPhotForcedCover}).

\begin{table}
	\fontsize{7}{8}\selectfont
	\caption{New flux densities for the 67 LERs categorized by C15.}
    \label{tab:Photometry1}
	\begin{tabular}{crrrrr} 
    	\hline
		C15 ID  & $S$\textsubscript{70$\mathrm{\mu}$m} & $S$\textsubscript{160$\mathrm{\mu}$m} & $S$\textsubscript{250$\mathrm{\mu}$m} & $S$\textsubscript{350$\mathrm{\mu}$m} & $S$\textsubscript{500$\mathrm{\mu}$m} \\
        & (Jy) & (Jy) & (Jy) & (Jy) & (Jy)  \\
		\hline
        2  &  206.11 &  319.40  &  220.12  &  87.38  &  26.79 \\ 
		6  &  267.46  &  189.83  &  66.60  &  24.5  &  6.11 \\ 
		7  &  59.01  &  54.7  &  22.82  &  9.48  &  2.55 \\ 
		10  &  105.28  &  85.76  &  43.88  &  17.25  &  5.37 \\ 
		11  &  214.43  &  416.36  &  268.18  &  128.42  &  42.16 \\ 
		16  &  181.17  &  186.23  &  103.43  &  44.08  &  17.79 \\ 
		18  &  202.9  &  444.37  &  273.72  &  118.48  &  38.12 \\ 
		19  &  790.69  &  830.54  &  922.33  &  386.54  &  126.92 \\ 
		20  &  678.05  &  404.12  &  153.49  &  52.76  &  16.27 \\ 
		22  &  151.43  &  175.25  &  125.41  &  51.02  &  16.92 \\ 
		23  &  114.17  &  144.16  &  82.19  &  37.13  &  11.13 \\ 
		26  &  193.28  &  321.65  &  35.62  &  14.37  &  4.00 \\ 
		32  &  1615.26  &  1016.05  &  362.23  &  120.41  &  38.97 \\ 
		33  &  2505.09  &  2229.98  &  1075.42  &  432.39  &  142.87 \\ 
		34  &  155.75  &  189.5  &  87.34  &  40.01  &  11.71 \\ 
		36  &  169.75  &  106.49  &  27.28  &  10.54  &  2.66 \\ 
		39  &  148.14  &  140.85  &  116.31  &  52.24  &  14.16 \\ 
		43  &  563.31  &  406.76  &  147.09  &  54.96  &  16.24 \\ 
		46  &  789.34  &  413.14  &  156.47  &  54.65  &  26.95 \\ 
		47  &  142.52  &  147.83  &  75.66  &  32.27  &  10.3 \\ 
		48  &  124.96  &  93.20  &  74.84  &  25.95  &  7.99 \\ 
		51  &  83.50  &  199.81  &  177.19  &  51.81  &  16.75 \\ 
		57  &  1089.21  &  926.63  &  375.85  &  143.81  &  46.02 \\ 
		58  &  58.96  &  33.41  &  8.62  &  4.93  &  2.47 \\ 
		59  &  550.74  &  373.71  &  173.86  &  60.27  &  17.62 \\ 
        60  &  207.57  &  311.23  &  148.46  &  62.09  &  20.75 \\ 
		61  &  270.35  &  392.31  &  190.96  &  74.13  &  22.76 \\ 
		62  &  102.3  &  146.52  &  53.96  &  16.94  &  3.28 \\ 
		65  &  174.78  &  268.82  &  201.62  &  75.36  &  23.62 \\ 
		70  &  706.48  &  627.68  &  700.19  &  243.38  &  65.03 \\ 
		74  &  227.02  &  278.17  &  121.26  &  48.45  &  14.45 \\ 
		87  &  192.46  &  91.054  &  39.64  &  14.4  &  5.36 \\ 
		90  &  332.45  &  321.79  &  91.35  &  39.85  &  15.94 \\ 
		92  &  212.24  &  335.59  &  195.15  &  90.83  &  32.66 \\ 
		94  &  415.35  &  318.45  &  273.93  &  111.49  &  35.38 \\ 
		96  &  609.70  &  421.63  &  138.31  &  48.78  &  13.21 \\ 
		98  &  102.19  &  86.32  &  32.91  &  11.89  &  2.94 \\ 
		99  &  89.28  &  75.45  &  27.23  &  10.82  &  3.11 \\ 
		101  &  358.69  &  550.59  &  340.87  &  141.74  &  47.10 \\ 
		102  &  226.93  &  163.77  &  41.86  &  13.55  &  3.80 \\ 
		105  &  198.03  &  156.5  &  77.49  &  30.26  &  9.43 \\ 
		115  &  542.21  &  32.52  &  92.79  &  27.89  &  7.24 \\ 
		116  &  156.68  &  252.29  &  181.23  &  79.49  &  27.29 \\ 
		122  &  177.07  &  300.83  &  236.12  &  143.54  &  49.44 \\ 
		123  &  670.20  &  662.87  &  512.94  &  212.79  &  69.89 \\ 
		124  &  272.03  &  192.85  &  93.19  &  35.29  &  10.94 \\ 
		129  &  282.12  &  128.11  &  41.05  &  13.27  &  3.99 \\ 
		131  &  207.22  &  159.71  &  59.87  &  23.53  &  7.70 \\ 
		142  &  40.37  &  83.39  &  39.48  &  19.61  &  7.02 \\ 
		149  &  113.49  &  102.81  &  41.80  &  16.73  &  5.62 \\ 
		150  &  612.32  &  764.32  &  473.31  &  196.12  &  64.69 \\ 
		152  &  255.65  &  283.53  &  149.38  &  59.66  &  18.48 \\ 
		154  &  30.51  &  26.48  &  12.85  &  5.94  &  1.90 \\ 
		155  &  99.71  &  101.93  &  67.69  &  24.00  &  7.40 \\ 
		158  &  309.60  &  232.64  &  60.49  &  21.90  &  15.68 \\ 
		160  &  272.50  &  223.53  &  81.94  &  31.06  &  10.30 \\ 
        161  &  418.13  &  506.15  &  231.19  &  99.41  &  32.19 \\ 
		163  &  343.41  &  22.40  &  77.67  &  27.56  &  8.17 \\ 
		175  &  660.28  &  629.34  &  603.98  &  261.18  &  88.84 \\ 
		183  &  2880.41  &  1843.9  &  578.84  &  191.38  &  57.31 \\ 
		184  &  202.05  &  133.06  &  43.90  &  15.44  &  5.09 \\ 
		186  &  1342.30  &  850.25  &  263.60  &  105.30  &  30.58 \\ 
		194  &  279.07  &  254.47  &  77.27  &  28.06  &  8.59 \\ 
		195  &  134.75  &  105.32  &  74.52  &  27.67  &  9.20 \\ 
		196  &  237.00  &  252.46  &  95.23  &  35.30  &  12.96 \\ 
		198  &  588.12  &  607.32  &  276.07  &  107.39  &  34.22 \\ 
		201  &  259.25  &  290.22  &  132.67  &  52.14  &  17.19 \\     
		\hline
    \end{tabular}
\end{table}

\subsection{Covering Factor} 
\label{sec:coverfact}

If the ionizing source of the \ion{H}{ii} region is not entirely enshrouded, photons are being lost to space rather than heating and ionizing the surrounding gas. This affects measurements in both the radio and infrared, which in turn influences the derived $N_\mathrm{Ly}$ and luminosity, respectively. We can quantify this by defining a geometric covering factor $f_c$. 

For a newly formed region it is reasonable to set $f_c \sim 1$. In this case the \ion{H}{ii} region would be ionization bounded, and the surrounding ISM would act as an effective bolometer. As a region evolves it is expected to clear out surrounding material through phenomena such as bipolar outflows \citep{yorke99}. The \ion{H}{ii} region can then become density bounded in certain directions and non-ionizing photons can also escape. The later stages of \ion{H}{ii} region evolution typically involve blister structures where very low covering factors, $f_c \leq 0.5$, would be appropriate.

\begin{figure}
	\includegraphics[width=\columnwidth]{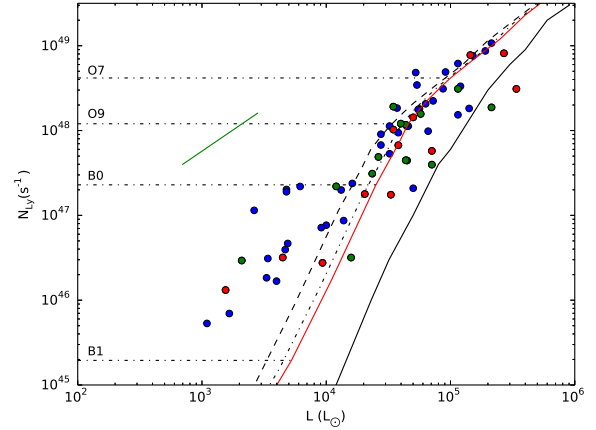}
    \caption{Lyman continuum flux of the 67 C15 LERs, as in \autoref{fig:Nly_l_ExcessMyPhot}, with recalculated luminosities from new photometry as well as newly scaled $N_\mathrm{Ly}$ from new kinematic distances. The dot-dash line corresponds to a $N_\mathrm{Ly}$-$L$ relationship for a ZAMS star with a $f_c = 0.8$, and the dashed line $f_c = 0.5$.}
    \label{fig:Nly_l_MyPhotForcedCover}
\end{figure}

We chose a reasonable covering factor for our sample, $f_c = 0.8$ as well as a representative low covering factor, $f_c = 0.5$, and applied these to the $N_\mathrm{Ly}$-$L$ calibration. The effect of a non-unity covering factor is to slide the calibration to lower $N_\mathrm{Ly}$ and luminosity along the direction indicated by the green line in \autoref{fig:Nly_l_MyPhotForcedCover}. Even when $f_c = 0.5$, there are still 20 high confidence sources that show Lyman excess. However, as stated above, using a covering factor of 0.5 would mean assuming a blister morphology for each source, which, from visual inspection of the sources, is not an accurate representation of the sample. For the purpose of eliminating sources from consideration, we will assume that $f_c = 0.8$ is possible. This results in $\sim 33\% - 50\%$ of the original LERs being removed from the sample (see \autoref{tab:distances} and \autoref{tab:photometry} for the exact numbers).  

We note our implicit assumption that the covering factor for both the ionized gas and the dust are equal does not necessarily need to be true. For example, an ionized-bounded \ion{H}{ii} region could be surrounded by a thin shell of material resulting in an ionized gas cover factor, $f_i \approx 1$, but a dust covering factor $f_d \ll 1$. In this case the surrounding ISM does not act as an effective bolometer, and points on the $N_\mathrm{Ly}$-$L$ plot would move to the right at constant $N_\mathrm{Ly}$. To explain the LER phenomenon in this manner though would typically require a very contrived molecular cloud morphology, which, once again, we see no evidence for in the images. 

\subsection{Different Stellar Calibrations}
\label{sec:model}

As C15 noted, the use of different $N_\mathrm{Ly}$-$L$ stellar calibrations \citep[e.g.,][]{panagia, crowther, davies} will change the number of \ion{H}{ii} regions classified as LERs. In general we find that the PAN73 calibration minimizes the Lyman excess over most of the luminosity range being considered; however, the \citet{davies} calibration does have a higher $N_\mathrm{Ly}$ than PAN73 for stars with log\textsubscript{10}($L$/$L$\textsubscript{\sun}) $>$ 4.75 while being comparable to PAN73 elsewhere.  Switching the $N_\mathrm{Ly}$-$L$ stellar calibration from PAN73 to \citet{davies} only results in a low number ($\sim 5$) of additional \ion{H}{ii} regions being removed from the LER sample (see \autoref{tab:distances} and \autoref{tab:photometry} for the exact numbers).

\section{Discussion} 
\label{sec:discussion}

\subsection{Physical Properties of the Sample}
\label{sec:physical}

\begin{figure}
	\includegraphics[width=\columnwidth]{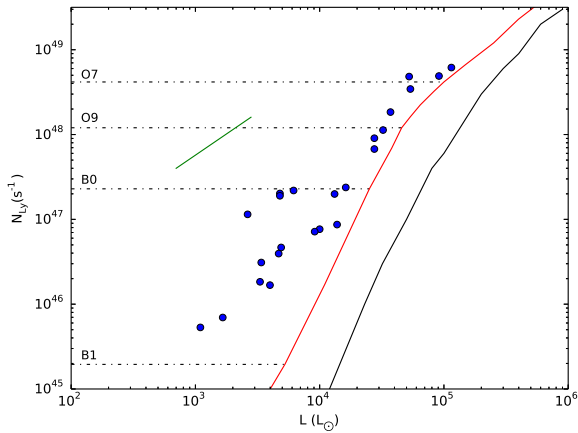}
    \caption{Lyman continuum flux versus luminosity as in \autoref{fig:Nly_l_MyPhotForcedCover} showing our refined subset of 24 LERs.}
    \label{fig:Nly_l_Gold}
\end{figure}

After applying the various modifications, described in \autoref{sec:analysis} to the original LER sample, we found that there was no reasonable way to move 24 of the sources out of the forbidden region of the $N_\mathrm{Ly}$-$L$ plot. This refined sample of LERs is summarized in \autoref{tab:gold_standards} and illustrated in \autoref{fig:Nly_l_Gold}. The quoted radius was determined from the \textit{Spitzer} $8.0~\micron$ images. It is the average of the distance from the centre of the source to where the intensity fell to $\sim5\%$ above the estimated background measured along lines of constant Galactic longitude and latitude. All of the sizes are comparable to typical sizes expected for photodissociation regions surrounding compact/ultra-compact \ion{H}{ii} regions \citep[e.g.,][]{kerton02,diaz-miller98}. 

This sample is not meant to be exhaustive; it is possible that some of the objects we removed from the original LER sample, and even some of the 133 C15 objects that fell in the allowed region of the $N_\mathrm{Ly}$-$L$ plot, are actually LERs. Rather, it represents an ideal sample for examining the LEP with a minimal amount of confusion from regular \ion{H}{ii} regions.

From \autoref{fig:Nly_l_Gold} we see that as the LERs decrease in luminosity the magnitude of the LEP increases. At $L\sim10^{5}L_\odot$ the excess in $N_\mathrm{Ly}$ is only a fraction of a dex, around $L=10^{3.5}-10^{4}L_\odot$ it increases to  approximately one dex, while at $L<10^{3.5}L_\odot$ the excess rises to two or more dex. The average $L$ and $N_\mathrm{Ly}$ of our final sample is $\log_{10}(L/L_\odot)=4.36$ and $\log_{10}(N_\mathrm{Ly}/\mathrm{s}^{-1}$) $=48.03$, whereas the averages of the sample that we removed are 4.89 and 48.30 respectively. This suggests that the phenomenon may be more prevalent in early B-type stars, although it is also possible that the phenomenon is just more easily detected in these stars compared with O-type stars.

\begin{figure}
	\includegraphics[width=\columnwidth]{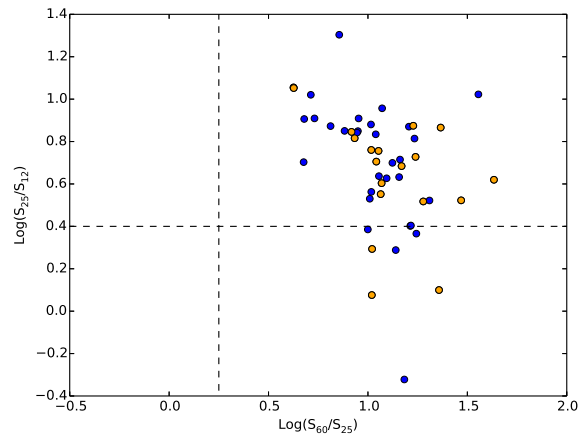}
    \caption{Colour-colour diagram using \emph{IRAS} flux densities for C15 excess regions which have associated \emph{IRAS} sources. Dashed lines show the devision between planetary nebulae (log(S\textsubscript{60}/S\textsubscript{25}) $< 0.25$, log(S\textsubscript{25}/S\textsubscript{12}) $> 0.4$)  and \ion{H}{ii} regions (log(S\textsubscript{60}/S\textsubscript{25}) $> 0.25$, log(S\textsubscript{25}/S\textsubscript{12}) $> 0.4$) \citep{Hughes89}. Dots in orange correspond to regions in the LER sample.}
    \label{fig:colour}
\end{figure}

\autoref{fig:colour} shows a colour-colour plot for C15 excess regions with coincident \emph{IRAS} point sources. Here we define coincidence as a separation $<2'$ between \emph{IRAS} and CORNISH point sources; however, 47 of the 49 coincident sources have a separation $<50''$. The close positional match was used to select sources that are likely the dominant source within the low-resolution ($1'-4'$) \emph{IRAS} beam thus minimizing confusion effects. Both our LERs and the C15 excess regions fall in the region expected for \ion{H}{ii} regions, as determined by \citet{Hughes89}, and there is clearly no contamination from planetary nebulae. 

\begin{figure}
	\includegraphics[width=\columnwidth]{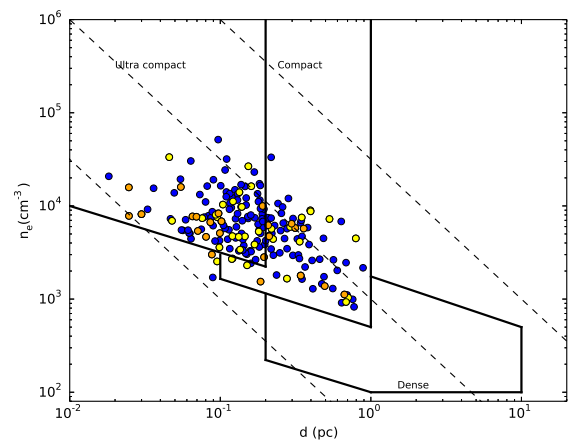}
    \caption{Electron density vs. 5 GHz diameter for all C15 regions using diameters from the CORNISH catalogue. Orange dots denote our refined sample of LERs, yellow the C15 regions we no longer consider excess, and blue the remaining C15 regions. The distinction between ultra-compact, compact, and dense regions are defined in \citet{Habing}. The dashed lines correspond to lines of constant excitation parameter of 10, 100, and 1000 pc cm\textsuperscript{-2} from left to right.}
    \label{fig:5GHzSize}
\end{figure}

As expected, given the selection criteria of the CORNISH survey, a plot of of electron density ($n_e$) versus physical (5 GHz) diameter ($d$) for all of the C15 sources (\autoref{fig:5GHzSize}) shows the sources falling into the canonical ultra-compact and compact bins designated by \citet{Habing}. The $n_e$ distribution of the two samples appears to be essentially identical, while the $d$ distribution of the 24 LERs appears to be slightly shifted to smaller sizes.

A Kolmogorov-Smirnov (KS) test was applied to the $n_e$ and $d$ distributions of the 24 LERs and the remaining 176 CORNISH regions. For $n_e$ the p-value was 0.40 indicating no statistically significant difference between the distributions. In contrast, a p-value of 0.008 was returned for $d$, suggesting the 24 LER sources are from a different (physically smaller) distribution. However, we caution that this may be a selection effect caused by our removal of sources from consideration due to potential for confusion. More compact \ion{H}{ii} regions were more likely to be classified as being isolated and thus considered to have higher quality photometry.

\begin{table*}
	\centering
	\caption{The refined sample of LERs. Values for luminosity, ionizing photon flux, and distance ($d$) reflect our new photometry and new kinematic distances. Radius ($r$) was determined using \textit{Spitzer} 8~\micron~images. The spectral index ($\alpha$) was calculated from data available in the literature with the error associated with the data fit shown in parentheses.}
	\label{tab:gold_standards}
	\begin{tabular}{ccccccccccc} 
		\hline
		C15  & CORNISH  & $l$  & $b$ & log\textsubscript{10}($L$/$L$\textsubscript{\sun}) & log\textsubscript{10}($N_\mathrm{Ly}/\mathrm{s}^{-1}$) & $d$ & r & $n$\textsubscript{$e$}  & SpT & $\alpha$ \\
          ID & ID & (deg.) & (deg.) & &  & (kpc) & (pc) & (10\textsuperscript{3} cm\textsuperscript{-3}) & ZAMS\textsuperscript{a} &  \\
		\hline         
		2 & G010.3204-00.2586 & 10.3204 & -0.2586 & 3.60 & 46.22 & 3.05 & 0.07 & 15.8 & B1 & -\\
		6 & G010.9656+00.0089 & 10.9656 & +0.0089 & 3.69 & 46.67 & 3.03 & 0.38 & 5.36 & B0 & -0.83 (.05)\\
		7 & G011.0328+00.0274 & 11.0328 & +0.0274 & 3.04 & 45.73 & 3.05 & 0.10 & 7.35 & B1 & 0.36 (.15)\\
        10 & G011.1712-00.0662 & 11.1712 & -0.0662 & 4.57 & 48.26 & 13.48 & 0.84 & 1.09 & O9 & 0.80 (.05)\\
        11 & G011.9032-00.1407 & 11.9032 & -0.1407 & 3.67 & 46.60 & 3.1 & 0.11 & 8.98 & B1 & 0.09 (.07)\\
        18 & G012.9995-00.3583 & 12.9995 & -0.3583 & 3.22 & 45.84 & 1.83 & 0.08 & 8.52 & B1 & -0.27 (.08)\\
        20 & G013.3850+00.0684 & 13.3850 & +0.0684 & 3.68 & 47.31 & 1.84 & 0.16 & 3.01 & B0 & 0.87 (.09)\\
        47 & G019.7281-00.1135 & 19.7281 & -0.1135 & 3.53 & 46.49 & 3.46 & 0.21 & 6.24 & B1 & -0.16 (.06)\\
        48 & G019.7407+00.2821 & 19.7407 & +0.2821 & 4.72 & 48.68 & 14.22 & 1.85 & 0.75 & <O7 & -0.22 (.04)\\
        51 & G020.3633-00.0136 & 20.3633 & -0.0136 & 3.42 & 47.06 & 3.44 & 0.14 & 11.1 & B0 & -0.09 (.06)\\
        61 & G023.2654+00.0765 & 23.2654 & +0.0765 & 4.12 & 47.30 & 4.77 & 0.26 & 6.12 & B0 & 0.07 (.05)\\
        74 & G024.9237+00.0777 & 24.9237 & +0.0777 & 3.68 & 47.28 & 3.33 & 0.22 & 1.74 & B0 & -0.48 (.05)\\
        92 & G027.5637+00.0845 & 27.5637 & +0.0845 & 4.51 & 48.05 & 8.37 & 0.61 & 1.43 & O9 & -0.16 (.04)\\
        94 & G027.9782+00.0789 & 27.9782 & +0.0789 & 4.21 & 47.38 & 4.41 & 0.31 & 2.54 & B0 & -0.17 (.03)\\
        102 & G028.6869+00.1770 & 28.6869 & +0.1770 & 4.44 & 47.83 & 8.07 & 0.48 & 4.11 & O9 & 0.07 (.09)\\
        124 & G032.2730-00.2258 & 32.2730 & -0.2258 & 4.96 & 48.69 & 12.6 & 1.05 & 2.25 & <O7 & 0.02 (.03)\\ 
        142 & G035.4570-00.1791 & 35.4570 & -0.1791 & 4.00 & 46.88 & 10.11 & 0.28 & 5.53 & B0 & -0.33 (.09)\\
        154 & G037.9723-00.0965 & 37.9723 & -0.0965 & 3.79 & 47.34 & 10.24 & 0.48 & 5.20 & B0 & 0.33 (.1)\\
        155 & G038.6465-00.2260 & 38.6465 & -0.2260 & 3.52 & 46.26 & 4.0 & 0.22 & 5.96 & B1 & 0.01 (.06)\\
        158 & G038.8756+00.3080 & 38.8756 & +0.3080 & 5.06 & 48.79 & 14.2 & 0.68 & 9.70 & <O7 & 0.31 (.04)\\
        194 & G052.7533+00.3340 & 52.7533 & +0.3340 & 4.73 & 48.54 & 9.5 & 0.82 & 3.49 & O8 & -0.38 (.03)\\
        195 & G053.1865+00.2085 & 53.1865 & +0.2085 & 4.44 & 47.96 & 9.75 & 0.82 & 1.58 & O9 & -0.05 (.04)\\
        196 & G053.9589+00.0320 & 53.9589 & +0.0320 & 3.96 & 46.86 & 3.95 & 0.15 & 13.5 & B0 & 0.11 (.05)\\
        198 & G059.6027+00.9118 & 59.6027 & +0.9118 & 4.14 & 46.94 & 3.57 & 0.03 & 27.7 & B0 & 0.73 (.07)\\
        \hline
        \multicolumn{10}{l}{\footnotesize$^\mathrm{a}$Spectral type consistent with $N_\mathrm{Ly}$ from \citet{panagia}.}\\ 
	\end{tabular}
\end{table*}

\begin{figure}

	\includegraphics[width=\columnwidth]{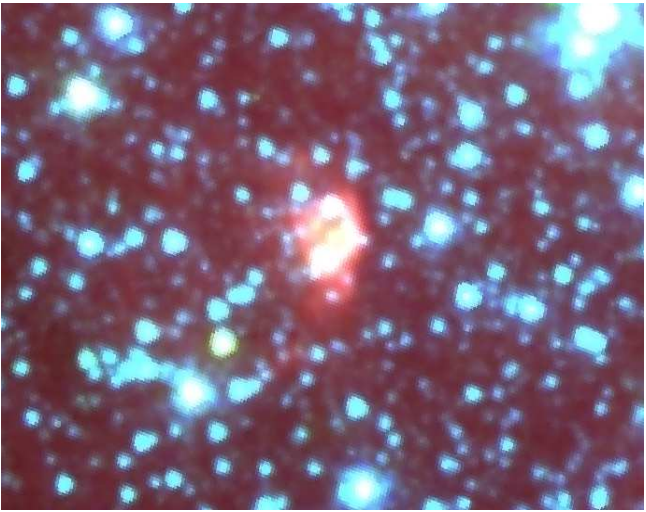}
    \caption{\textit{Spitzer} $2.75'$ x $2.1'$ false colour (RGB) image (8.0, 4.5, 3.6 $\mathrm{\mu}$m) centred on the coordinates from Table \ref{tab:gold_standards} for source 10.}
    \label{fig:a}
\end{figure}

\begin{figure}
	\includegraphics[width=\columnwidth]{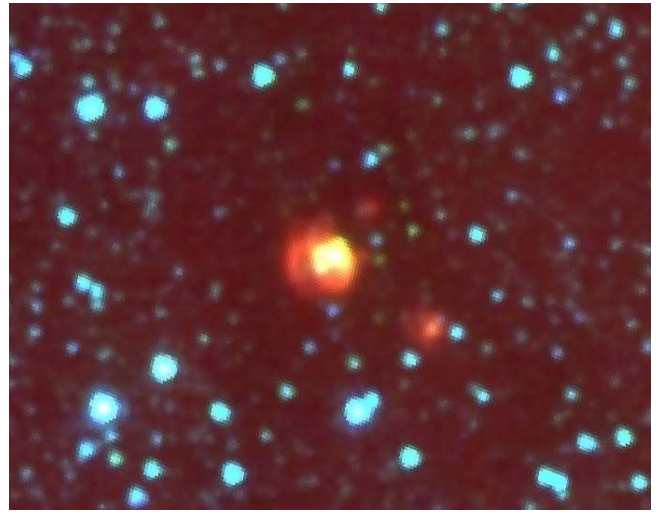}
    \caption{As in \autoref{fig:a} for source 18.}
    \label{fig:b}
\end{figure}

\begin{figure}
	\includegraphics[width=\columnwidth]{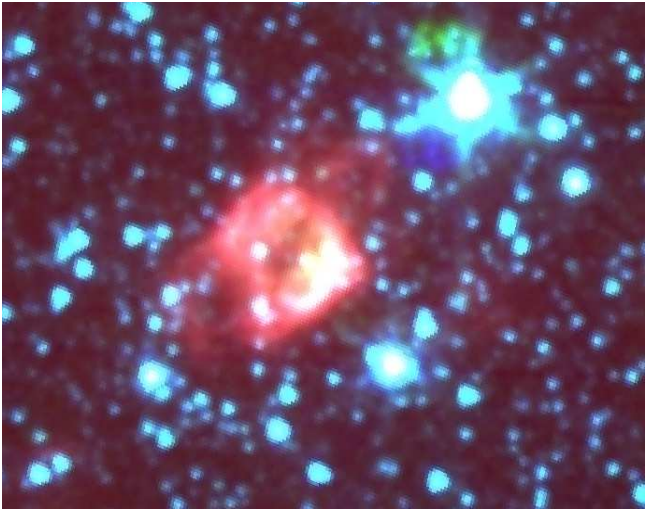}
    \caption{As in \autoref{fig:a} for source 48.}
    \label{fig:c}
\end{figure}

\begin{figure}
	\includegraphics[width=\columnwidth]{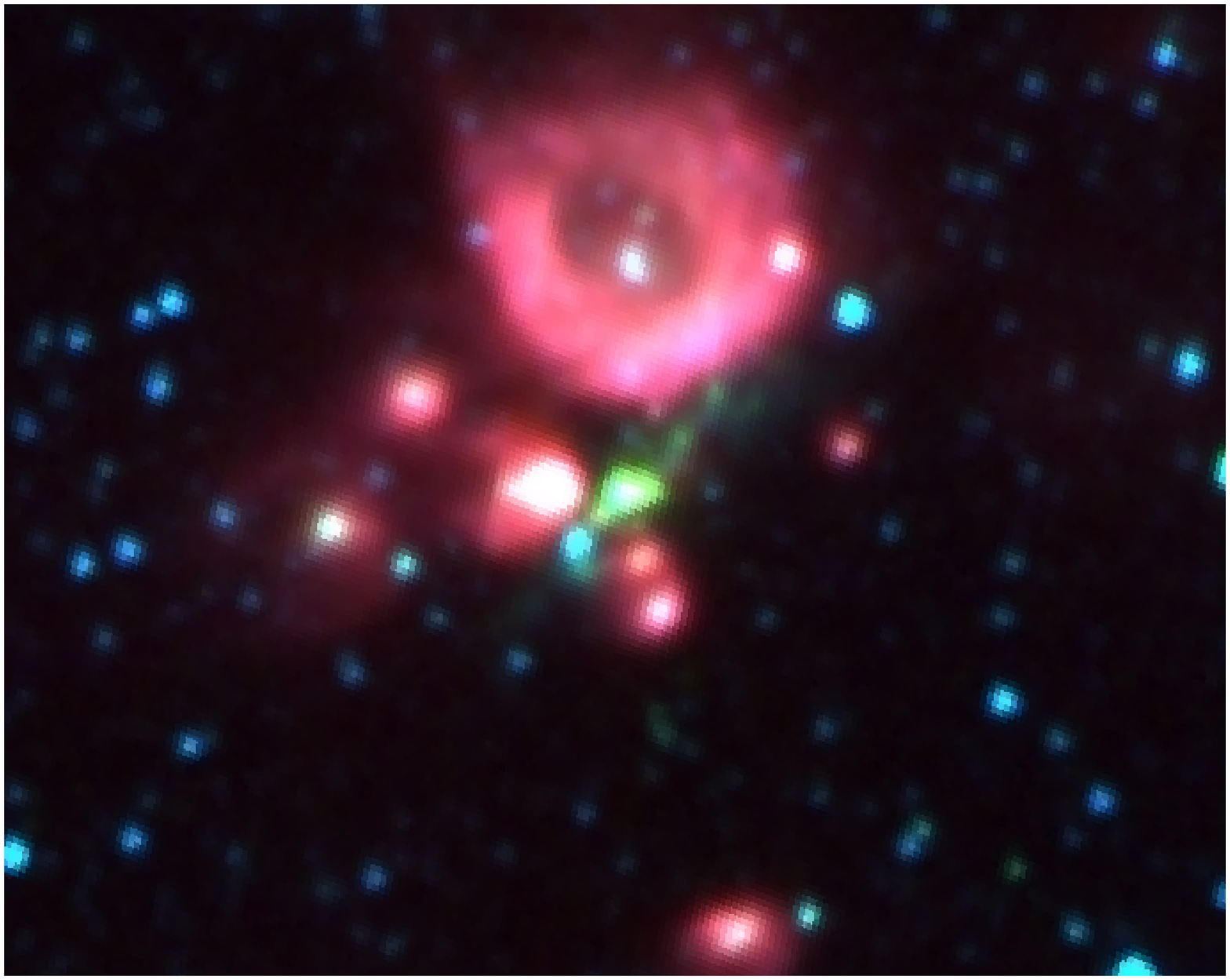}
    \caption{As in \autoref{fig:a} for source 198. This source is discussed in more detail in \autoref{sec:discussion}.}
    \label{fig:d}
\end{figure}

Inspection of \textit{Spitzer} three-colour (3.6, 4.5, and 8.0 \micron) images of our 24 sources show a variety of morphologies, but in general there is nothing about their infrared morphology that makes this sample of regions stand out from the other CORNISH \ion{H}{ii} regions (see examples in \autoref{fig:a}, \autoref{fig:b}, and \autoref{fig:c}). However, in our inspection we did find one peculiar \ion{H}{II} region, source 198 (\autoref{fig:d}; Marshall et al. in preparation), that shows very bright extended emission at 4.5 \micron. Such emission is often associated with very young outflow activity \citep{ego,grace}. In such cases it is possible that the observed \ion{H}{II} region is actually an ionized portion of the outflow (see further discussion in \autoref{sec:jets}).

The values of the basic physical properties, IR colour, density, and size, of our LER subsample all fall within the ranges expected for ultra-compact and compact \ion{H}{ii} regions. In the following sections we examine different possible origins for the LEP in these otherwise normal looking \ion{H}{ii} regions. In \autoref{sec:bac}, \autoref{sec:jets}, and \autoref{sec:disk} we assume that the underlying stellar models (e.g., PAN73 or \citet{davies}) are correct and examine other sources of ionizing photons besides the LyC. Then, in \autoref{sec:stars}, we discuss evidence that the LyC flux is being underestimated by stellar models, and explore the extent to which existing models need to be modified in order to match the LER observations.

\subsection{Ionization from the Balmer Continuum}
\label{sec:bac}

The LEP has been noted in a number of previous studies of young stellar objects \citep[YSOs;][]{simon81,Simon,Bally83}, molecular outflows \citep{snellandbally}, and outer Galaxy star-forming regions \citep{Mead}. In all of these cases, it was suggested that the additional ionization required originated from Balmer continuum (BaC) photons.

\begin{figure}
\includegraphics[width=\columnwidth]{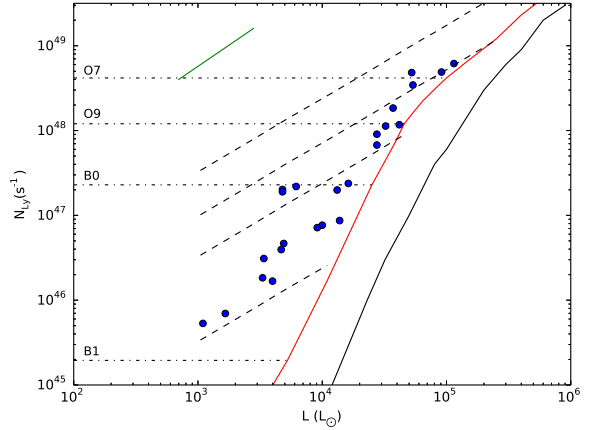}
    \caption{Lyman continuum flux versus luminosity as in \autoref{fig:Nly_l_Gold} with the dashed lines showing a range of 1, 10, 30 and 100\% of the Balmer continuum photons produced by blackbodies with the same T\textsubscript{eff} and luminosity as the PAN73 spectral types. Comparison with actual stellar models shows that there is no significant difference in the Balmer continuum flux from OB stars or equivalent blackbodies in the range of interest.}
    \label{fig:Nly_l_Gold_balmer}
\end{figure}

\begin{figure}
	\includegraphics[width=\columnwidth]{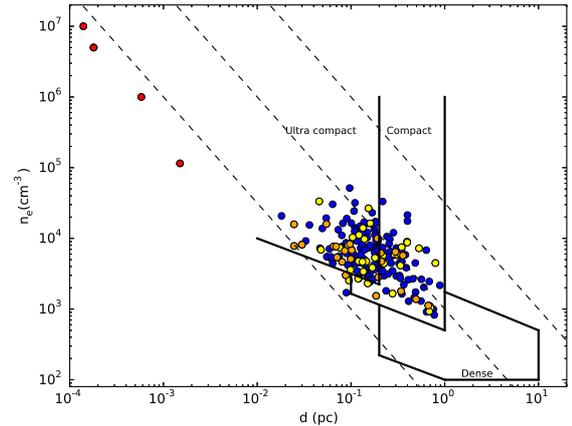}
    \caption{As in \autoref{fig:5GHzSize}, but note the scaling has been stretched to also display, in order of increasing electron density, NGC 2071-IRS 1 \citep{Bally83}, M8E, CRL 490, and the BN object \citep{simon81} in red.}
    \label{fig:5GHzSizeLARGE}
\end{figure}

Allowing ionization by BaC photons would easily solve the LER problem since $N\textsubscript{Ba} >> N_\mathrm{Ly}$ for all OB stellar models. As shown in \autoref{fig:Nly_l_Gold_balmer}, if only 10\% of the BaC photons that are produced actually help to ionize the region, the excess problem is solved for half of the LERs. BaC ionization appears to be important though only in high density ($10^5-10^7$ cm$^{-3}$), compact ($10^{-4}-10^{-2}$ pc) regions, where the n=2 level of hydrogen can be populated \citep{Simon,Thompson1984}. As shown in \autoref{fig:5GHzSizeLARGE}, the types of objects where BaC ionization is thought to play an important role, such as the Becklin-Neugebauer (BN) Object \citep{BN}, are 2 -- 4 orders of magnitude denser and smaller than the \ion{H}{II} regions we are considering. 

As part of their study of compact radio sources associated with molecular outflows, \citet{snellandbally} developed a radio spectral index and flux criteria to identify potential BaC ionized stellar winds. Such regions had a radio spectral index between $0.5<\alpha<1.5$ and an angular size related to its radio flux at 5 GHz by:
\begin{equation}
\label{eqn:size}
\theta_d \mathrm{(arcsec)} =  0.072 \left(\frac{S_\nu}{\mathrm{mJy}}\right). 
\end{equation}

We calculated radio spectral indices for our LER sources using data available in the literature (primarily from \citet{purcell,Ofek} and \citet{Giveon}; see \autoref{tab:gold_standards}). Only three of the LERs meet the \citet{snellandbally} spectral index criteria, and their size at 5 GHz is far larger than that predicted from \autoref{eqn:size}. We note that some of the clearly spurious results in \autoref{tab:gold_standards} (e.g., $\alpha=-0.86$  for source 6) likely arise from a combination of having only a limited number of observations and/or insufficient frequency coverage. 

While ionization by BaC photons is a viable explanation for the intensity of the radio emission observed in compact, BN-type objects, we conclude that BaC ionization is not likely to be an important process in the more extended, lower density interstellar medium associated with the CORNISH \ion{H}{ii} regions.

\subsection{Shock Excitation in Winds and Jets}
\label{sec:jets}

Outflows associated with lower luminosity (solar-type) YSOs have been observed having radio continuum emission 2-13 orders of magnitude larger than would be expected from a plasma ionized by the stellar LyC flux \citep{anglada1996}. Because of the extended nature of these objects, \citet{anglada1996} concluded that BaC ionization does not play a key role, rather they suggest that shocks in a neutral stellar wind are the cause of the excess ionization. 

\begin{figure}
\includegraphics[width=\columnwidth]{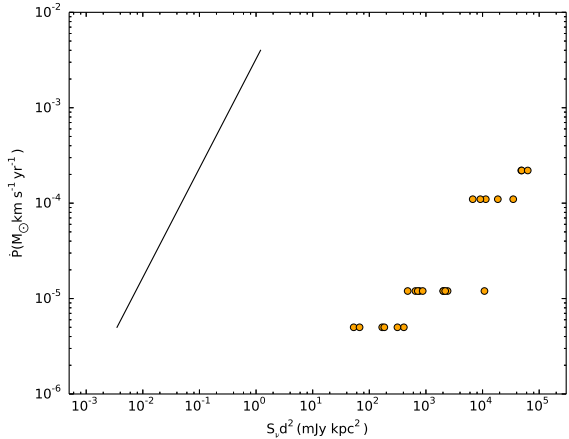}
    \caption{Momentum rate vs. radio continuum luminosity of LERs. Momentum rates were determined using characteristic terminal velocities and mass-loss rates from \citet{Muijres,krticka} based on the spectral types given in \autoref{tab:gold_standards}. The line shows the trend found in \citet{anglada1996} for radio jets associated with lower luminosity stars.}
    \label{fig:Anglada}
\end{figure}

To determine if this process could be applicable to our LERs we calculated momentum rates using terminal velocities and mass-loss rates for OB stars described in \citet{Muijres,krticka} and plotted them against the radio continuum luminosity of the objects (see \autoref{fig:Anglada}). We find that, while our calculated momentum rates for LERs fall within the same range as their sample ($\sim 10^{-5}-10^{-2} M_\odot$ km s$^{-1}$ yr$^{-1}$) the LERs are roughly 5 dex more radio luminous.

HH 80 and 81 are Herbig-Haro (HH) objects believed to be powered by the high-luminosity YSO {\it IRAS}~18162-2048. The size and electron density of both HH 80 and 81 are such that they could be easily mistaken for traditional (i.e., containing a massive star) compact \ion{H}{II} regions based solely on the size and intensity of their 5 GHz emission. We find that $N_\mathrm{Ly} \sim 10^{44}$~s$^{-1}$ for HH 80 and 81, comparable to a B2-B3 ZAMS star. While this is non-negligible, it is still approximately an order a magnitude lower in $N_\mathrm{Ly}$ than our weakest LER. In addition, the infrared luminosity of these objects is many orders of magnitude lower than the range of bolometric luminosities we are investigating. 

We conclude that shock ionization processes like those associated with outflows/winds from lower-luminosity YSOs are not responsible for the LER phenomenon. Also confusion between traditional \ion{H}{II} regions and HH objects is unlikely.

\subsection{Accretion Shock Emission}
\label{sec:disk}

Excess optical and ultraviolet emission from young, solar-mass YSOs (i.e., T Tauri stars) has long been explained as being due to emission associated with shocks forming as material accretes onto the forming star. The original versions of these models had the emission originating in a boundary layer where material would transition from orbital velocities to stellar rotation velocities over a very short distance \citep{lynden,bbb}. More recent models have material being channeled along magnetic field lines from an accretion disk to the stellar photosphere with the emission occurring as the material shocks on impact with the photosphere \citep{koenigl, hartmann94}.

Disregarding the fact that the magnetospheric channeling mechanism is questionable for OB-type stars \citep{Hartmann2016}, we can construct a simple model for this process, following \citet{matsuyama}, where the additional accretion shock emission is modelled as a blackbody. For the blackbody temperature we used 15,000 K, which is almost certainly an overestimate \citep{Hartmann98, matsuyama}. The accretion shock luminosity $L_\mathrm{as}$ was calculated using:
\begin{equation}
\label{eqn:las}
L_\mathrm{as} =  \frac{GM_{*}\dot{M}_d}{2R_{*}}  ,
\end{equation}
where,  $\dot{M}_d$ is the accretion rate, and $M_{*}$ and $R_{*}$ are the stellar mass and radius respectively. Using representative values of mass and radius for early B-type/late O-type stars ($\sim10$~M$_\odot$, $\sim5$~R$_\odot$), and a very high accretion rate $10^{-4}$~M$_\odot$yr$^{-1}$, \autoref{eqn:las} gives a luminosity of $\sim10^{3.5}$~L$_\odot$. 

The {\sc{cloudy}} photoionization code \citep{cloudy17} was then used to combine TLUSTY NLTE stellar atmosphere models \citep{TLUSTY1,TLUSTY2} corresponding to B2~V, B1~V, and O9~V stars (using the \citet{conti08} calibration) with the blackbody representing the additional accretion shock emission. The results are shown in \autoref{tab:models} in the TLUSTY+BB rows. In the lowest luminosity model the addition of the blackbody does add 1.2 dex to $\log(Q)$, but this is still short of the value required to match our LER observations (e.g., compare with the $N_\mathrm{Ly}$ and $L$ values shown in \autoref{fig:Nly_l_Gold_balmer}). At higher luminosities, the model for the accretion shock emission had essentially no effect on the overall ionizing photon flux.

Even with the generous blackbody temperature and $\dot{M}$ values chosen, the TLUSTY models with the additional blackbody emission fall short in producing the ionizing photon flux required to power the LERs.  We conclude that the additional ionizing flux in the LERs does not originate from a scaled-up version of the disk accretion process around low-mass YSOs.

\subsection{Enhanced Lyman Continuum Stellar Models}
\label{sec:stars}

Observations of the LyC emission from OB stars are extremely challenging due to the large distances to the nearest such stars combined with strong interstellar absorption at these frequencies. Direct evidence that the LyC flux from OB stars is being underestimated by current stellar atmosphere models comes primarily from extreme ultraviolet (EUV) observations of the B2 II star $\epsilon$~CMa and the B1 II-III star $\beta$~CMa obtained by the {\it EUV Explorer} ({\it EUVE}) satellite \citep{Cass,Cass1996}. These two stars are relatively nearby, 188 pc and 206 pc respectively, and lie along a relatively low density line of sight in the local interstellar medium \citep{PowerLaw2}. 

As discussed more fully in both \citet{vac96} and \citet{PowerLaw2}, {\it EUVE} observations of $\epsilon$~CMa indicate an $N_\mathrm{Ly}$ excess of a  factor of 30 (1.5 dex) \citep{Cass}. The star's low mass loss rate, and the existence of a mid-infrared excess at 12 and 25 $\mu$m  support the idea that the excess EUV emission arises in the stellar photosphere, possibly due to an increase in temperature due to wind blanketing \citep{PowerLaw2}. The interpretation of the $\beta$~CMa observations are not as clear. \citet{Cass1996} derive two models of the star with different effective temperatures; the lower temperature model yields an excess of a factor of 5 (0.7 dex) while the higher temperature model has essentially no excess emission. 

Motivated by these observations, we wanted to determine if slight modifications of stellar models could match our LER observations. Using the {\sc{cloudy}} photoionization code \citep{cloudy17} we modified the SEDs of our TLUSTY models from \autoref{sec:disk} in the EUV and X-ray regimes in two different ways. 

\begin{figure}
\subfloat[B2~V Model]{%
  \includegraphics[clip,width=\columnwidth]{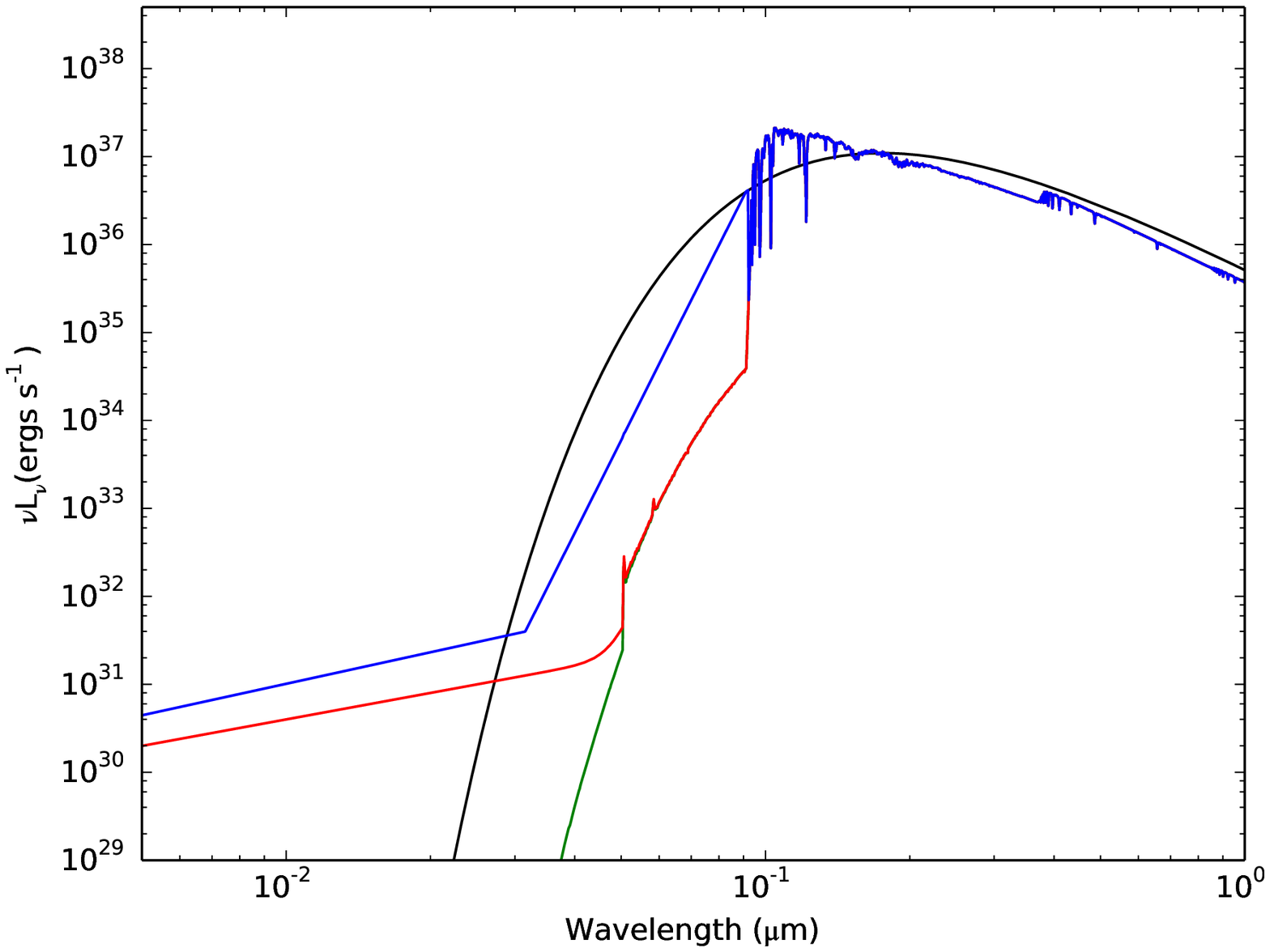}%
}

\subfloat[B1~V Model]{%
  \includegraphics[clip,width=\columnwidth]{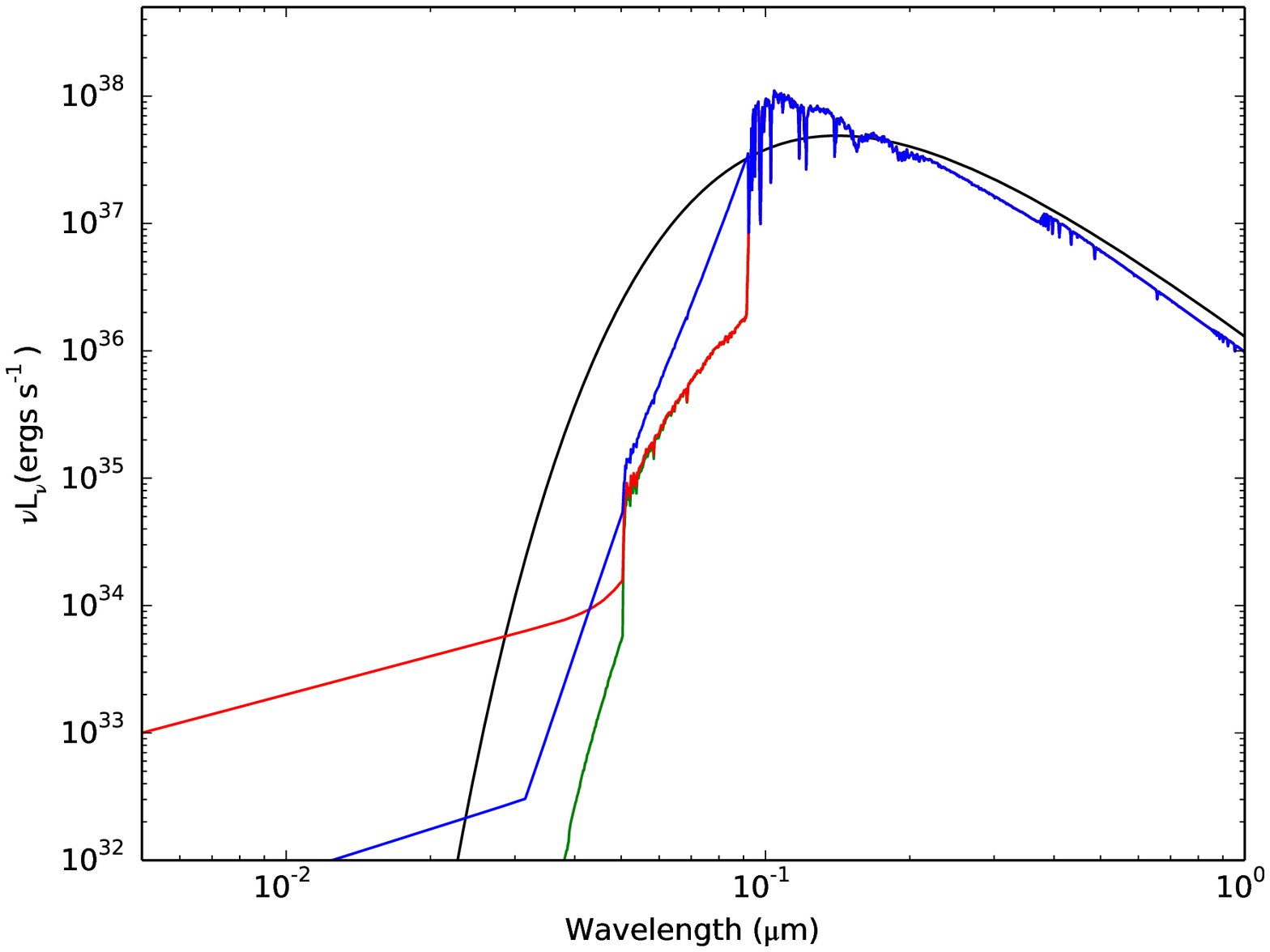}%
}

\subfloat[O9~V Model]{%
  \includegraphics[clip,width=\columnwidth]{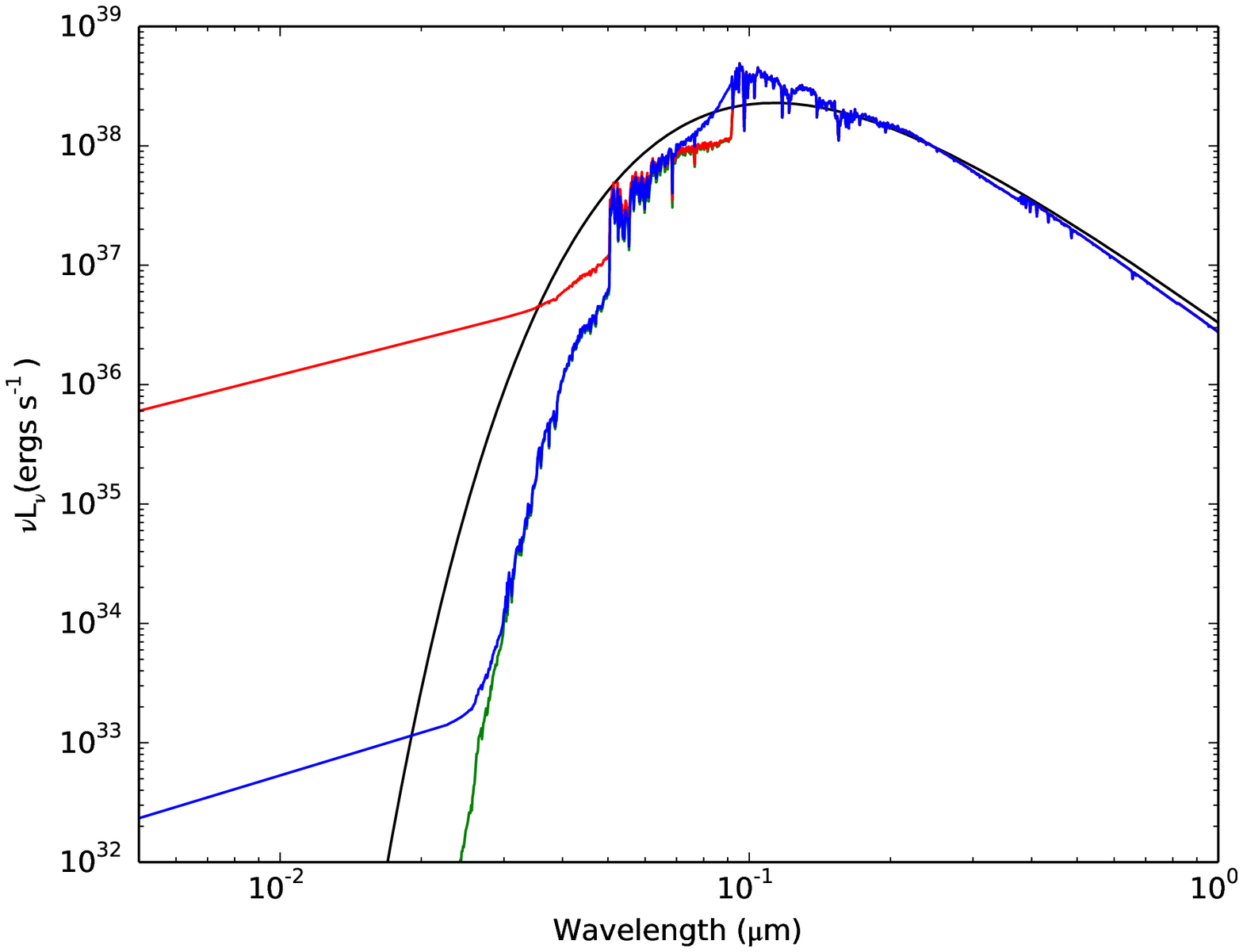}%
}
\caption{SEDs of blackbodies and TLUSTY stellar atmospheres. Temperatures and model data are found in \autoref{tab:models}. Lines in black correspond to blackbody emission. Green lines correspond to the unmodified TLUSTY models with the red and blue power law modifications over-plotted. Red corresponds to the power law modeled after \citet{PowerLaw1}, and blue corresponds to the piecewise SED modeled after \citet{PowerLaw2,Cass}. Spectral type equivalents are from the \citet{conti08} calibration.}
\label{fig:Models}
\end{figure}

\citet{PowerLaw1} found that the addition of shocks to a stellar atmosphere model produced additional EUV (shortward of 500~\AA) and X-ray emission. We matched their model SEDs (see their Figure 1) by the addition of a power law spectrum, $F_\nu \propto \nu^{-2}$ between $\sim$46 and 600~\AA, normalized to the TLUSTY model at 500~\AA. The modified SEDs are shown in \autoref{fig:Models}, and the results of this modification on the LyC flux are shown in \autoref{tab:models} in the TLUSTY+PL1 rows. We find that the effect on the LyC flux is minimal at all luminosities. While this sort of power law may be useful in explaining higher energy X-ray emission from OB stars, it does not come close to matching the LER observations.

We also modified the TLUSTY models to match the EUV emission seen in the spectrum of $\epsilon$ CMa \citep[see Figure 6.2 in][]{PowerLaw2} by adding a piecewise power law, normalized to the TLUSTY model at 912~\AA, to the TLUSTY SED. The power law has a $F_\nu \propto \nu^{-2}$ dependency between $\sim$45 and 315~\AA\ followed by a very steep increase to 912~\AA. The modified SEDs are shown in \autoref{fig:Models}, and the results of this modification on the LyC flux are shown in \autoref{tab:models} in the TLUSTY+PL2 rows. This modification does give the increased ionization flux that we see in LERs, and it also reproduces the reduction in the excess observed as we move from B-type to O-type stars. 

\subsection{A Young Star Phenomenon?}
\label{sec:young}

Our analysis of the CORNISH \ion{H}{II} regions shows that, even though direct observations of the LEP are limited to only two evolved giant and bright giant stars, the phenomenon is not restricted to these more evolved stages of stellar evolution. In fact there is some evidence that the phenomenon may be preferentially associated with younger OB stars. 

In \autoref{sec:physical} we showed that it is possible that our LERs are preferentially smaller, and thus likely to be younger, than the non-LER \ion{H}{II} regions. Of course, the size of an \ion{H}{II} region is only a rough proxy for its age due to variations in ISM structure, and there is also some concern that the difference is a selection effect. However, there is additional support for the idea that the LEP is preferentially associated with younger \ion{H}{ii} regions from a different study.

\citet{Sanchez} also describe LERs in their study of southern-sky ultra-compact and compact \ion{H}{II} regions. They subdivided their \ion{H}{II} region sample into three types depending on their association with millimeter and infrared emission using this as a proxy for age: Type 1 sources only had associated millimeter emission, Type 2 had both millimeter and infrared emission, and Type 3 had only infrared emission. They found that the excess is observed predominantly in Type 1 and 2 sources and suggest, as we do, that the LEP is more common in younger, more compact \ion{H}{II} regions. They also speculate that the excess UV radiation originates from a protostar/disk system, but our analysis in \autoref{sec:disk} does not support this idea.

\begin{figure}
\includegraphics[width=\columnwidth]{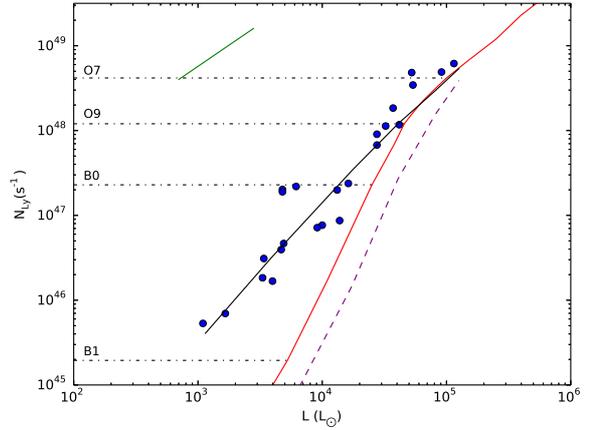}
    \caption{Lyman continuum flux versus luminosity as in \autoref{fig:Nly_l_Gold}. Solid black curve shows the blackbody calculated using calibration from \citet{conti08} and CLOUDY modeling to obtain the ionizing flux for a blackbody of the same temperature and luminosity. The red line showing the PAN73 calibration and the dashed purple line showing the TLUSTY NLTE, line-blanketed models. }
    \label{fig:Nly_l_Gold_Blackbody}
\end{figure}

Many of the excess sources in \citet{Sanchez} and in our study lie very close to the blackbody emission line in the $N_\mathrm{Ly}$-$L$ plot (\autoref{fig:Nly_l_Gold_Blackbody}). An extreme alternative to the interpretation presented in the previous three paragraphs is that this locus essentially represents the correct location of all main-sequence OB stars in the $N_\mathrm{Ly}$-$L$ plot. If that is the case though, the allowed region, defined by clusters of stars, must also be shifted over to lower luminosities. For early O-type stars the stellar models already converge toward the blackbody line, so the biggest change would be for \ion{H}{II} regions powered by B-type stars. Regions that were once explained as containing a B-star along with a cluster of lower mass stars would now become highly overluminous. This is not a trivial issue considering that 80\% of the \ion{H}{II} regions consistent with B-type $N_\mathrm{Ly}$ can be explained in terms of a traditional stellar model combined with various degrees of clustering. Since direct EUV observations of the stars powering these \ion{H}{II} regions are impossible, determining the extent of the LEP in OB stars will rely primarily on future studies of the stellar content/clustering found in the most luminous B-star powered \ion{H}{II} regions.   

\begin{table}
	\centering
	\caption{Description of TLUSTY models with the various SED additions.}
	\label{tab:models}
	\begin{tabular}{lcccc} 
		\hline        
		Model (Z = Z\textsubscript{\sun}, & T\textsubscript{(eff)} & log\textsubscript{10}($L$/$L$\textsubscript{\sun}) & log(Q) & $\Delta$log(Q) \\
        log(g) = 4.0) & (kK) &&&\\
		\hline
        TLUSTY & 21 & 3.59 & 44.34 & - \\
        TLUSTY + PL1\textsuperscript{a} & 21 & 3.59 & 44.34 & 0.00 \\
        TLUSTY + PL2\textsuperscript{b} & 21 & 3.60 & 46.20 & 1.86 \\
        TLUSTY + BB\textsuperscript{c} & 21 & 3.85 & 45.54 &  1.20 \\
		\hline         
        TLUSTY & 26 & 4.24 & 46.17 & - \\
        TLUSTY + PL1 & 26 & 4.24 & 46.18 & 0.01 \\
        TLUSTY + PL2 & 26 & 4.26 & 47.12 & 0.95 \\
        TLUSTY + BB & 26 & 4.31 & 46.25 & 0.09 \\ 
        \hline
        TLUSTY & 32.5 & 4.91 & 48.19 & - \\
        TLUSTY + PL1 & 32.5 & 4.91 & 48.23 & 0.04 \\
        TLUSTY + PL2 & 32.5 & 4.94 & 48.37 & 0.18 \\
        TLUSTY + BB & 32.5 & 4.93 & 48.19 & 0.00 \\
        \hline
        \multicolumn{5}{p{\linewidth}}{$^\mathrm{a}$TLUSTY modified with the power law addition modeled after stellar atmosphere shocks \citep{PowerLaw1}.}\\
        \multicolumn{5}{p{\linewidth}}{$^\mathrm{b}$TLUSTY modified with the power law addition modeled after EUV observations of $\epsilon$ CMa \citep{PowerLaw2,Cass}.}\\
         \multicolumn{5}{p{\linewidth}}{$^\mathrm{c}$TLUSTY with the addition of a 15~kK blackbody to simulate accretion shock ionization as described in \autoref{sec:disk}.}\\
        \end{tabular}        
\end{table}

\section{Conclusions} 
\label{sec:conclusions}

We began with an initial sample of 67 sources categorized as compact and ultra-compact \ion{H}{II} regions in the CORNISH catalogue and further classified as LERs in C15. The goal of our study was to refine the initial sample by removing LERs that could be otherwise explained by observational uncertainties or more conventional astrophysical processes. We examined the effect of: 1) recalculating luminosities; 2) obtaining new distance estimates; 3) performing new photometry; 4) applying reasonable covering factors; and 5) considering different stellar calibrations. We have clearly shown that the LER phenomenon cannot be explained away through this combination of factors, and our final refined sample of LERs consists of 24 sources. The general properties of the refined sample, such as IR colour, PDR size, and electron density, are indistinguishable from those expected for regular compact and ultra-compact \ion{H}{ii} regions. 

While this phenomenon has already been observed around different types of objects, we find that the solutions typically proposed in the literature, Balmer continuum ionization, emission from wind/jet shocks, and emission associated with accretion shocks, do not adequately explain the magnitude of the effect in the CORNISH \ion{H}{II} regions. In addition they do not explain the trends we see with spectral type: two-thirds of our final LER sample are associated with B-type stars, and the magnitude of the Lyman excess increases as you move from O-type stars to B-type stars.

We conclude that the most likely cause of the LEP in the majority of the CORNISH \ion{H}{II} regions is a modification to the underlying photosphere, perhaps ultimately caused by wind line blanketing, resulting in increased EUV emission especially just shortward of 912\AA. 

On balance, we prefer the picture where the LEP is primarily associated with the OB stars powering the youngest \ion{H}{II} regions, rather than being associated with all OB stars. Future studies of the stellar content of luminous B-star powered \ion{H}{II} regions will help determine if this is the case.

\bibliographystyle{mnras} 
\bibliography{Lyman_Excess}
\bsp	

\appendix
\section{Removed Sources}

\begin{table*}
	\centering
	\caption{Sources removed from the original LER sample. Luminosity, photon flux, and distance values are based on our new photometry and kinematic distance.}
	\label{tab:non-excess}
	\begin{tabular}{cccccccc}
		\hline
		C15 \# & CORNISH & log\textsubscript{10}($L$/$L$\textsubscript{\sun}) & log\textsubscript{10}($N$\textsubscript{$Ly$}/$s$\textsuperscript{-1}) & $d$  &  $n$\textsubscript{$e$}  & Notes\textsuperscript{a}\\
         ID   &  ID &   &  & (kpc) & (10\textsuperscript{3} cm\textsuperscript{-3}) & & \\
		\hline
		16 & G012.4294-00.0479 & 5.15 & 48.26 & 20.19 & 4.66 & P17,P15,S,C\\
		19 & G013.2099-00.1428 & 4.54 & 48.28 & 4.55 & 8.35 & PC\\
        22 & G014.1741+00.0245 & 4.82 & 47.99 & 14.41 & 1.39 & P17,P15,L15,C\\
		23 & G014.4894+00.0194 & 3.32 & 46.47 & 2.89 & 4.99 & PC\\
        26 & G016.3913-00.1383 & 4.75 & 48.26 & 12.13 & 1.11 & C\\
        32 & G018.1460-00.2839 & 4.54 & 48.01 & 3.5 & 1.93 & PC\\
        33 & G018.3024-00.3910 & 5.06 & 48.19 & 3.5 & 4.77 & P17,P15,L15,C\\
        34 & G018.4433-00.0056 & 4.64 & 48.07 & 12.01 & 9.71 & C,PC\\
        36 & G018.6738-00.2363 & 4.74 & 48.25 & 12.76 & 1.03 & C\\
        39 & G018.8250-00.4675 & 3.97 & 46.44 & 4.93 & 5.28 & PC \\
        43 & G019.4912+00.1352 & 5.43 & 48.91 & 14.2 & 1.84 & P17,C,PC\\
        46 & G019.6781-00.1318 & 5.33 & 48.27 & 11.98 & 0.93 & P17,P15,C\\
        57 & G021.4257-00.5417 & 4.70 & 47.32 & 4.76 & 1.91 & P17,P15,C\\
        58 & G021.6034-00.1685 & 4.42 & 47.69 & 15.71 & 2.17 & L17,L15,PC\\
        59 & G021.8751+00.0075 & 5.33 & 49.03 & 13.87, & 0.22 & D\\
        60 & G023.1974-00.0006 & 4.20 & 46.50 & 5.7& 5.38 & P17,C,PC\\
        62 & G023.4553-00.2010 & 3.85 & 46.69 & 5.87 & 9.78 & PC \\
        65 & G023.8618-00.1250 & 4.65 & 47.65 & 10.7& 3.29 & P17,P15,C,PC\\
        70 & G024.4736+00.4950 & 4.64 & 47.65 & 5.9 & 4.91 & P17,P15,C,PC\\
        87 & G026.6089-00.2121 & 5.18 & 48.87 & 19.5 & 2.69 & L17,D\\
        90 & G027.2800+00.1447 & 5.16 & 48.89 & 13.5 & 5.72 & PC \\
        96 & G028.2447+00.0131 & 4.85 & 47.60 & 8.0 & 2.00 & P17,P15,C,PC\\
        98 & G028.4518+00.0027 & 4.08 & 47.34 & 8.11 & 8.07 & P17,S,C,PC\\
        99 & G028.5816+00.1447 & 4.58 & 47.97 & 15.4 & 1.84 & L17,L15,C,PC\\
        101 & G028.6523+00.0273 & 4.70 & 48.16 & 8.0 & 5.56 & P17,P15,C\\
        105 & G030.2527+00.0540 & 4.65 & 48.05 & 10.8 & 2.98 & P17,L15,C\\
        115 & G031.0709+00.0508 & 5.08 & 48.52 & 11.7 & 22.6 & P17,P15,C\\
        116 & G031.1596+00.0448 & 3.19 & 46.12 & 2.05 & 24.7 & PC\\
        122 & G032.0297+00.0491 & 4.52 & 47.24 & 8.1 & 2.87 & P17,P15,C,PC\\
        123 & G032.1502+00.1329 & 4.76 & 48.20 & 5.4 & 3.18 & P17,P15,C\\
        129 & G032.9273+00.6060 & 5.28 & 48.94 & 17.5 & 3.18 & P17,C\\
        131 & G033.4163-00.0036 & 4.58 & 47.83 & 9.48 & 1.51 & P17,C,PC\\ 
        149 & G037.7562+00.5605 & 4.51 & 47.73 & 12.3 & 1.28 & L17,L15,C\\
        150 & G037.7633-00.2167 & 5.06 & 48.49 & 9.65 & 1.49 & P17,P15,C,PC\\
        152 & G037.8683-00.6008 & 4.80 & 48.32 & 10.0 & 5.44 & P17,P15,C\\
        160 & G039.7277-00.3973 & 4.66 & 48.05 & 9.3 & 0.83 & P17,C\\
        161 & G039.8824-00.3460 & 4.86 & 48.35 & 9.0 & 10.4 & P17,P15,C\\
        163 & G041.7419+00.0973 & 4.94 & 48.49 & 11.7 & 4.63 & P17,P15,C\\
        175 & G043.8894-00.7840 & 4.31 & 47.25 & 3.38 & 14.0 & P,L15,C,PC\\
        183 & G045.4790+00.1294 & 5.53 & 48.49 & 7.92 & 2.19 & P17,P15,C,PC\\
        184 & G045.5431-00.0073 & 4.37 & 47.49 & 7.96 & 2.38 & PC\\
        186 & G048.9296-00.2793 & 4.85 & 47.76 & 5.6 & 5.26 & P17,P15,C,PC\\
		201 & G061.2875-00.3327 & 4.60 & 48.08 & 8.67 & 0.64 & PC\\
        \hline
        \multicolumn{7}{l}{\footnotesize$^a$ Indicates the factor(s) which recategorized the source as no longer a LERs.}\\ 
        \multicolumn{7}{l}{\hspace{5mm}P17 is using new photometry at the MK17 distance.}\\ 
        \multicolumn{7}{l}{\hspace{5mm}P15 is using new photometry at the C15 distance.}\\ 
        \multicolumn{7}{l}{\hspace{5mm}S is scaling the distance only.}\\ 
        \multicolumn{7}{l}{\hspace{5mm}C is using new photometry at the MK17 distance and a f\textsubscript{c} = 0.8}\\ 
        \multicolumn{7}{l}{\hspace{5mm}PC is reflective of a medium/low photometry confidence level.}\\
        \multicolumn{7}{l}{\hspace{5mm}L15 is using new luminosity calculated using C15 flux densities with C15 distances.}\\ 
        \multicolumn{7}{l}{\hspace{5mm}L17 is using new luminosity calculated using C15 flux densities with MK17 distances.}\\
        \multicolumn{7}{l}{\hspace{5mm}D is using the ZAMS stellar parameters from \citet{davies} only.}\\
	\end{tabular}
\end{table*}

\label{lastpage}
\end{document}